\documentclass[runningheads,a4paper]{llncs}

\usepackage{amssymb}
\usepackage{graphicx, subfigure}
\usepackage{amsmath}

\usepackage{url}
\urldef{\mailsa}\path|{Ehsan.Fattahi, Christian.Waluga, Barbara.Wohlmuth}@ma.tum.de|
\urldef{\mailsb}\path|Ulrich.Ruede@fau.de|

\newcommand{\keywords}[1]{\par\addvspace\baselineskip
\noindent\keywordname\enspace\ignorespaces#1}
\newcommand{\vect}[1]{\mathbf{#1}}
\newcommand{\citep}[1]{\cite{#1}}

\newcommand{\waLBerla}{\textsc{waLBerla}}
                                                                                                                                     
\begin{document}

\mainmatter  

\title{Large scale lattice Boltzmann simulation for the coupling of free and porous media flow}
\titlerunning{LBMs for free and porous media flow}

%
%
\author{Ehsan Fattahi \and Christian Waluga\and Barbara Wohlmuth\and Ulrich R\"ude}
\authorrunning{Lecture Notes in Computer Science: Authors' Instructions}

\institute{Technische Universit\"at M\"unchen\\
Fakult\"at f\"ur Mathematik\\
\mailsa\\
\mailsb\\
\url{http://www-m2.ma.tum.de}}

\toctitle{Lecture Notes in Computer Science}
\tocauthor{Authors' Instructions}
\maketitle

\begin{abstract}

In this work, we investigate the interaction of free and porous media flow by large scale lattice Boltzmann simulations. We study the transport phenomena at the porous interface on multiple scales, i.e., we consider both, computationally generated pore-scale geometries and homogenized models at a macroscopic scale. The pore-scale results are compared to those obtained by using different transmission models. Two-domain approaches with sharp interface conditions, e.g., of Beavers--Joseph--Saffman type, as well as a single-domain approach with a porosity depending viscosity are taken into account. For the pore-scale simulations, we use a highly scalable communication-reducing scheme with a robust second order boundary handling. We comment on computational aspects of the pore-scale simulation and on how to generate pore-scale geometries. The two-domain approaches depend sensitively on the choice of the exact position of the interface, whereas a well-designed single-domain approach can significantly better recover the averaged pore-scale results.


\end{abstract}

\section{Introduction}
Transport phenomena in porous materials are important in many scientific and engineering applications such as catalysis, hydrology, tissue engineering and enhanced oil recovery. In the past several decades, flow in porous media has been studied extensively both experimentally and theoretically. We refer the interested reader to the textbook \cite{Helmig:2011} and the references therein. In porous media flow, we usually distinguish between three scales: the pore-scale, the representative elementary volume (REV) scale and the domain scale. The REV is defined as the minimum element for which macroscopic characteristics of a porous flow can be observed. Because experimental setups for many practical questions may be too expensive or even impossible to realize, numerical simulation of porous media flow can be a useful complementary method to conventional experiments. 

To describe the flow in the bulk of the porous medium, {Darcy}'s law is commonly used in the form
\begin{equation}
	\label{eq:Darcy}
	\mu \vect{K}^{-1} \vect{u} = \vect{F} - \nabla p ,
\end{equation}
where $\mu$ is the dynamic viscosity of the fluid, $\vect{K}$ is the permeability tensor of the porous medium, $\vect{F}$ is the body force, and $\vect{u}$ and $p$ are averaged velocity and pressure quantities, respectively. However, when a porous medium and a free flow domain co-exist, e.g., in a river bed, there is no uniquely accepted model for the transition between the Darcy model and the free flow. Different approaches based on two-domain or on single-domain models are available. Using a single-domain in combination with the Brinkman equation that modifies {Darcy}'s law by a viscous term
\begin{equation}
	\label{eq:Brinkman}\tag{Br}
	-\mu_{\rm eff} \nabla^2 \vect{u} + \mu \vect{K}^{-1} \vect{u} = \vect{F} - \nabla p,
\end{equation}
allows to model a smooth transition (see e.g. \cite{Alazmi:2001,Nield:2009,Lebars:2006}). Here $\mu_{\rm eff}$ is an effective dynamic viscosity in the porous region. However, determining appropriate viscosity parameters for the Brinkman model in the transient region is challenging \citep{Lebars:2006,Goyeau:2003,Chandesris:2009}. Furthermore, the penetration of the flow into the porous medium is found to depend on the roughness coefficient of the surface; see e.g. \cite{Goharzadeh:2005,Ghisalberti:2010,Morad:2009,Pokrajac:2009}.

Alternatively, one can use a two-domain approach in combination with a sharp transmission condition.
Considering the (Navier-)Stokes equation in the free flow region and the Brinkman (or Darcy) equation in the porous region, the interface plays an important role. Proceeding from the experimental investigation of Poiseuille flow over a porous medium, Beavers and Joseph \cite{Beavers:1967} introduced an empirical approach that agreed well with their experiment; see also \cite{Nield:2009}: They suggested to use a slip-flow condition at the interface, i.e., the velocity gradient on the fluid side of the interface is proportional to the slip velocity. For simplicity, we consider a domain for which the interface is aligned with the flow direction. The Beavers--Joseph relation is formulated as
\begin{equation}
	\label{eq:BJ}\tag{BJ}
	\left.\frac{dU}{dz}\right|_{z=0^+} = \frac{\alpha}{\sqrt{k}} \left(U_s - U_m \right),
\end{equation}
where $z$ denotes the coordinate perpendicular to the interface, $U = U(z)$ is the mean velocity in flow direction, $U_s$ is the slip velocity at the interface $z = 0^+$, $U_m$ is the seepage velocity that is evaluated far from the plane $z=0$ in the porous region, $k$ is the permeability, and $\alpha$ is a phenomenological dimensionless parameter, only depending on the porous media properties that characterize the structure of the permeable material within the boundary region which typically varies between 0.01 and 5 \citep{Nield:2006,Duman:2009}. We refer to \citep{Baber:2012,Mosthaf:2011} and the references therein for the interface coupling of two-phase compositional porous-media flow and one-phase compositional free flow.

In 1971, Saffman \citep{Saffman:1971} found that the tangential interface velocity is proportional to the shear stress. He proposed a modification of the BJ condition as
\begin{equation}
	\label{eq:BJS}\tag{BJS}
	\frac{\sqrt{k}}{\alpha} \left.\frac{dU}{dz}\right|_{z=0^+} = \left(U_s - U_m \right) + O(k).
\end{equation}
More than two decades later, Ochoa-Tapia and Whitaker \cite{OchoaTapia:1995} proposed an alternative modification of the BJ condition
which includes the velocity gradient on both sides of the interface as
\begin{equation}
	\label{eq:WTK}\tag{OTW}
	\left. \mu_{\rm eff} \frac{dU}{dz}\right|_{z=0^-} - \left. \mu \frac{dU}{dz}\right|_{z=0^+} = \frac{\mu}{\sqrt{k}} \beta U_{s}.
\end{equation}
Here the jump-coefficient $\beta$ is a free fitting parameter that needs to be determined experimentally \citep{Martys:1994}. Different expressions for the effective viscosity $\mu_{\rm eff}$ can be found in the literature. For instance, Lundgren \cite{Lundgren:1972} suggested a relation of the form $\mu_{\rm eff} = \mu / \epsilon$. 

All of the interface conditions mentioned above require the a priory knowledge of the exact position of the interface \citep{Zhang:2009,Nabovati:2009,LIU:2011}, which is for realistic porous geometries often not the case. Additionally both, single-domain and two-domain, homogenized models rely on assumptions whose validity is not automatically guaranteed and depend on additional parameters. Traditional experiments to validate and calibrate such models are often costly, time consuming and difficult to set up. On the other hand, modern high performance computers enable the development of increasingly complex and accurate computational models resolving pore-scale features. Designing highly efficient solvers for partial differential equations is one of the challenges of extreme scale computing. While finite volume/element/difference schemes give rise to huge algebraic systems, lattice Boltzmann methods are intrinsically parallel and extremely attractive from the computational complexity point of view. Thus fully resolved direct numerical simulation based on first principles modeling is not only feasible nowadays but also provides an attractive possibility for validation and calibration. As a step in this direction, we here carry out a pore-scale simulation of free flow over a porous medium. The model porous media geometry is constructed by generating a random sphere-packing using an in-house multi-body simulation framework called PE \cite{Preclik:2015}. In the pore geometries constructed such, the flow equations are solved with full geometrical resolution.
This naturally leads to high computational cost requiring the use of high end parallel computing. As we will show by performance analysis, the in-house lattice Boltzmann solver \waLBerla{} \citep{feichtinger:2009} exhibits excellent performance and parallel scalability for these pore-scale simulations.

We use the results of the direct numerical simulation of flow over and through the porous media as reference solution and evaluate several sharp-interface conditions. As a further example, we also use a homogenized lattice Boltzmann model as a REV scale simulation and show the capability of this model to reproduce the pore-scale results with high accuracy.

\section{Numerical method}

The lattice Boltzmann method (LBM) has been successfully applied to simulate porous media flow. 
The kinetic nature of the LBM enables it for fluid systems involving microscopic interactions, e.g., flow through porous media. Furthermore, its computational simplicity, its amenability to simple and efficient implementation and parallelization, and its ability of handling geometrically complex domains make it an applicable tool to simulate porous media flow on the pore-scale \citep{Succi:1989,Singh:2000,Bernsdorf:2000,Kim:2001}.

The LBM can also be applied to model the fluid flow in porous media at the REV scale. The most commonly used models are the Darcy, the Brinkman-extended Darcy and the Forchheimer-extended Darcy models. This last approach accounts for the flow resistance in the standard LBM by modifying body-force or equilibrium terms, leading to the recovery of either Darcy-Brinkman’s equations or generalized Navier-Stokes equations \citep{Spaid:1997,Freed:1998,Martys:2001}.
The general model of porous media flow should consider the fluid forces and the solid drag force in the momentum equation \citep{Nithiarasu:1997}. Guo and Zhao \cite{Guo:2002} proposed a model to include the porosity into the equilibrium distribution and added a force term to the evolution equation to account for drag forces of the medium. The non-linear inertial term is not included in the Brinkman model either, and thus, is only suitable for low-speed flow. 
In this approach, the detailed structure of the medium is ignored, and the statistical properties of the medium are included directly. 

\subsection{The lattice Boltzmann equation}
The LBM originates from the lattice-gas automata method 
and can also be viewed as a special discrete scheme for the Boltzmann equation with discrete velocities 
\begin{equation}
	\label{eq:LBM}
	\vect{f}(\vect{x}+\vect{e}_k\Delta t , t +\Delta t ) -\vect{f}(\vect{x}, t) = \vect\Omega(\vect{x}, t) + F_k \Delta t
\end{equation}
where $\vec{e}_k$ is the particle velocity and $\vect\Omega(\vect{x}, t)$ is the collision operator.
For the three dimensional lattice model $D_3Q_{19}$,
$\vect{f}(\vect{x}, t) = ({f}_0(\vect{x}, t),{f}_1(\vect{x}, t),...,{f}_{18}(\vect{x}, t))^T$
is a 19 dimensional vector of distribution function.
$F_K$ is the force that acts as a source term to drive the flow.

It is very common to use the Bhatnagar-Gross-Krook (BGK)\citep{Bhatnagar:1954} model that features a single-relaxation-time (SRT) approximation for the collision operator. However, it has been shown that using the SRT leads to nonphysical viscosity dependence of boundary locations and also suffers from poor stability properties \citep{Pan:2006,Bogner2015}. Here, we use the TRT collision operator in which the relaxation time of the symmetric and anti-symmetric components of the distribution function are separated.
For an in-depth discussion of the TRT model we refer to \citep{Ginzburg:2007,Ginzburg:2008:a,Ginzburg:2008:b}.
As proposed by Ginzburg \cite{Ginzburg:2007}, the TRT model uses two relaxation rates $\omega^+$ and $\omega^-$ where $\omega^+$ is used for even order 
moments, and $\omega^-$ is used for odd order moments
\begin{equation}
	\label{eq:TRT}
	\vect\Omega(\vect{x}, t) = - \omega^+ \left(\vect{f}^+(\vect{x}, t) -
	\vect{f}^{eq,+}(\vect{x}, t) \right) - \omega^- 	
	\left(\vect{f}^-(\vect{x}, t) - \vect{f}^{eq,-}(\vect{x}, t) \right), 
\end{equation}
and 
\begin{equation}
	\label{eq:TRT2}
	f_k^+= \frac{f_k + f_{\bar{k}}}{2}  \  \ , \  \ f_k^-= \frac{f_k - f_{\bar{k}}}{2} .
\end{equation}
Here $\bar{k}$ is the index of the discrete velocity opposite to the one associated with the index $k$. The first eigenvalue is set to $1/\omega^+= 3 \nu+0.5$ and the second eigenvalue $\omega^-$ is a free parameter. Due to stability reasons, $ \omega^-$ has to be selected in $(0,2)$.
The equilibrium distribution function $\vect{f}^{\rm eq}(\vect{x}, t_n)$ for
incompressible flow is given by \citep{He:1997}
\begin{equation}
	\label{eq:equilibrium}
	{f}_k^{\rm eq}(\vect{x}, t_n) = w_k \left\lbrace \delta \rho + \rho_0 \left[ c_s^{-2}\vect{e}_k \cdot \vect{u} +\tfrac12 c_s^{-4}(\vect{e}_k \cdot 	\vect{u})^2 - \tfrac12 c_s^{-2}\vect{u} \cdot \vect{u} \right] \right\rbrace,
\end{equation}
\noindent where $w_k$ is a set of weights normalized to unity,
$\rho = \rho_0 + \delta \rho $. 
Here $\delta \rho $
is the density fluctuation, and $\rho_0$ is the mean density
which we set to $\rho_0=1$. 
$c_s = \Delta x/(\sqrt{3} \Delta t)$ is the lattice speed of sound, while $\Delta x$ denotes the lattice cell width.
The macroscopic values of density $\rho$ and velocity $\vect{u}$ 
can be calculated from $\vect{f}$ as zeroth and first order moments with respect to the particle velocity, i.e.,
\begin{equation}
	\label{Macro}
	\rho = \sum\nolimits_{k=0}^{18} f_k, \qquad \vect{u} = \rho_0^{-1} \sum\nolimits_{k=0}^{18} \vect{e}_k f_k.
\end{equation}

In a lattice Boltzmann scheme, we typically split the computation into a collision and a streaming step 
that are given as
\begin{align}
\label{eq:collision}\tag{collision}
\tilde{f}_k(\vect{x}, t_n) - f_k(\vect{x}, t_n) &= \vect\Omega(\vect{x}, t) + F_k \Delta t ,\\
\label{eq:streaming}\tag{streaming}
f_k(\vect{x}+\vect{e}_k\Delta t , t_{n+1}) &= \tilde{f}_k(\vect{x}, t_n),
\end{align}
respectively, for $k=0,\dots,18$.
The execution order of these two steps is arbitrary and may vary from code to code for implementation reasons.

In addition, for linear steady flow, it has been demonstrated \citep{Ginzburg:2008:a} that most of the macroscopic errors/quantities of the TRT depend on $\Lambda = \left( \frac{1}{\omega^+} - \frac{1}{2}\right)\left( \frac{1}{\omega^-} - \frac{1}{2}\right) $ the so-called magic parameter that includes the spatial error, stability, best advection and diffusion.
The choice $\Lambda = \frac{1}{4}$ is 
suggested as a suitable value for porous media simulations.
Another choice, namely $\Lambda = \frac{3}{16}$, yields the exact location of bounce-back
walls in case of Poiseuille flow in a straight channel \citep{Ginzburg:2008:a,Khirevich:2015}.
 
\subsection{Boundary conditions}
In this study, two types of boundary conditions are used for the pore-scale simulation. The first one is a no-slip wall condition and the second one is a periodic pressure forcing that is applied to drive the flow by a pressure gradient.
The simplest scheme to imply no-slip boundary conditions in lattice Boltzmann is the simple bounce-back (SBB) operator. In this scheme, the wall location is represented by a staircase approximation, and the no-slip boundary is satisfied by the bounce-back phenomenon of a particle reflecting its momentum upon collision with a wall. Hence, the unknown distribution function is calculated as:
\begin{equation}
	\label{eq:noslip}
	f_{\bar{k}}(x_{f_1}, t_{n+1})= \tilde{f}_k(x_{f_1},t_n).
\end{equation}
where we take the values $\widetilde{f}_k$ after collision but before streaming on the right hand side.
However, the staircase approximation is not appropriate for complex geometries
where more accurate results are required even for a low resolution of the boundary. 
Hence, the central linear interpolation (CLI) scheme which yields a higher accuracy at moderately increased computational cost is our preferred choice.

In the CLI scheme \citep{Ginzburg:2008:a} three particle distribution functions are
needed at two fluid nodes adjacent to the solid node, i.e.,
\begin{equation}
	\label{eq:CLI}
	f_{\bar{k}} (x_{f_1}, t_{n+1})= \tfrac{1-2q}{1+2q} \tilde{f}_k(x_{f_2},t_n) - \tfrac{1-2q}{1+2q} \tilde{f}_{\bar{k}}(x_{f_1},t_n) + \tilde{f}_k(x_{f_1},t_n).
\end{equation}
\noindent while $q=|x_{f_1}-x_w|/|x_{f_1}-x_b|$ defines a normalized distance of the first fluid node to the wall.
$x_{f_1}$ and $x_{f_2}$ are the first and second fluid neighbor cells in the direction of $\bar{k}$, respectively. We use the $\Lambda = \frac{3}{16}$ which the CLI has been shown to have second order accuracy \citep{Khirevich:2015}.

\section{Large scale simulations}
	\label{results}
In this study, we use the \waLBerla{} software framework \citep{feichtinger:2009,Feichtinger:2011} that provides a highly optimized implementation of the TRT model that is about as fast as the SRT model. We refer to \cite{Godenschwager:2013}, where scalability of \waLBerla{} to more than $10^{12}$ lattice cells and almost 500\,000 cores has been demonstrated.
Here, different from the SBB boundary condition, we use the CLI scheme that must access two neighboring fluid cells. In \waLBerla{}, this situation is handled by extra ghost-layer exchanges, i.e., by communicating an extended set of distribution functions to neighboring processors. This results in an additional communication in case of massively parallel simulation runs. 

\begin{figure}
\centering
\includegraphics[trim= 3mm 2mm 3mm 2mm,clip,width=.45\textwidth]{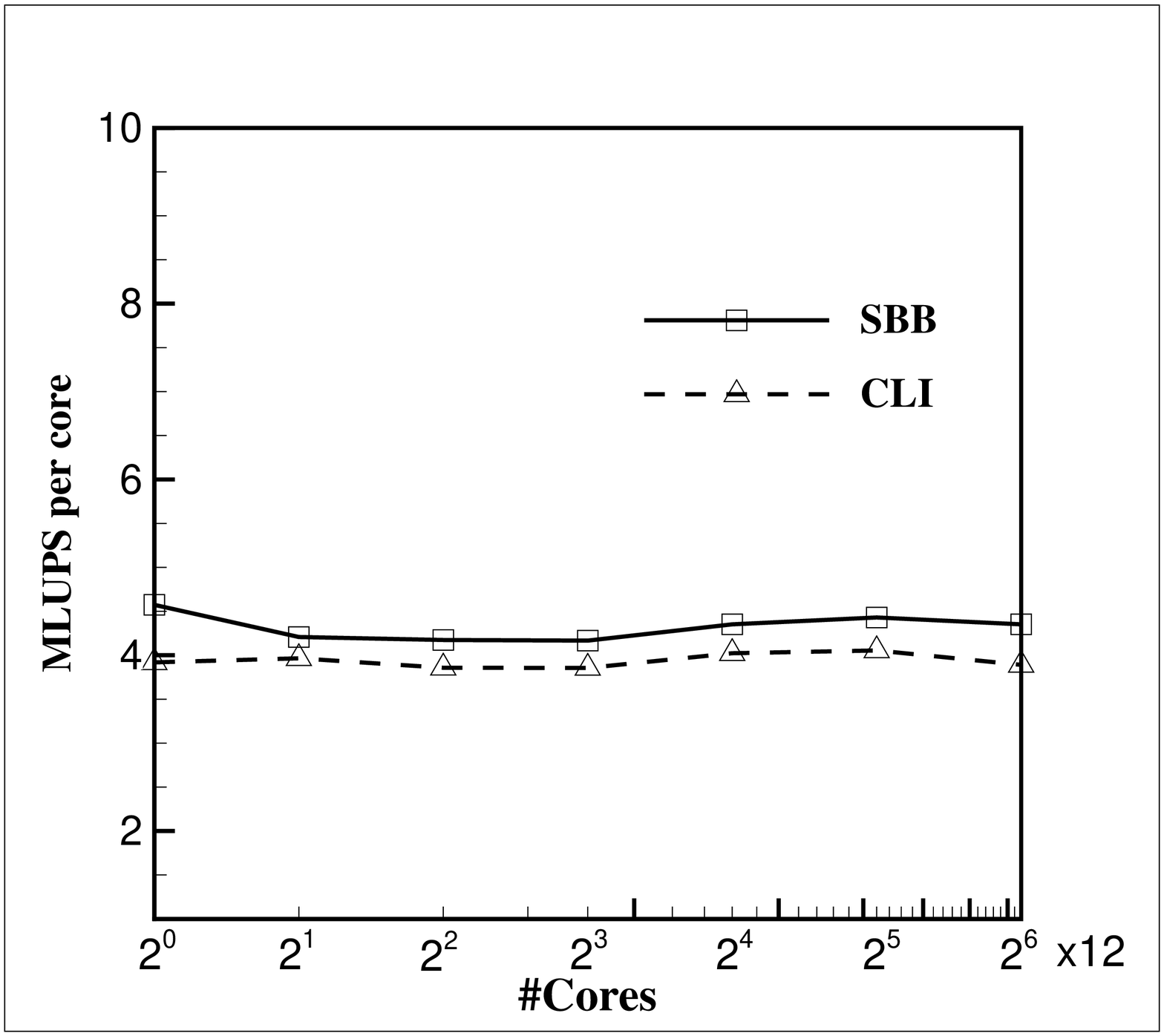}
  \caption{Measured MLUPS per core for Weak scaling on LIMA-Cluster using $151^3$ cells per core.
   }
  \label{fig:weakMLUPS}
\end{figure}

To demonstrate the parallel scalability and efficiency of the \waLBerla{} framework in the context of a porous media simulation, we first perform a weak-scaling study. Here we use a lattice of $151^3$ cells per core and embed into this grid a sphere with a diameter of 90 cells. The results have been obtained on the LIMA cluster at RRZE \footnote{https://www.rrze.fau.de/dienste/arbeiten-rechnen/hpc/systeme} which has 500 compute nodes. Each node consists of two Intel Xeon 5650 "Westmere" chips so that each node has 12 cores running at 2.66 GHz. 
We conduct scalability tests ranging from one node to 64 nodes. This setup results in $2.64\times10^9$ cells for the largest run including 768 spherical obstacles. Fig.~\ref{fig:weakMLUPS} displays the weak-scaling results using the TRT kernel. Fig.~\ref{fig:weakMLUPS} shows the mega lattice updates per second (MLUPS) for the SBB and CLI boundary schemes. The results do not only confirm that the code scales very well, but also that the MLUPS value per core compares favorably with many other LBM implementations \cite{Peters:2010,Krafczyk:2011,Robertsen:2015}.

We point out that achieving a good scaling behavior becomes more challenging when the node performance is already high, but that a high performance on each node is a fundamental (though sometimes neglected) prerequisite for achieving good overall performance. Thanks to both, the meticulously optimized \waLBerla{} kernels on each node, combined with the carefully designed communication routines, the MLUPS value per core is high and stays nearly constant while the number of cores is increased.
Note that the CLI boundary condition causes a slowdown of about 10\% in comparison to the SBB boundary condition, which is the fastest scheme. The slowdown of the performance while using the CLI is due to the additional time that is needed for the communication and the higher complexity of the boundary condition compared to SBB. However the higher accuracy of the CLI \citep{Fattahi:2015} compared to the SBB allows in complex application to use a coarser resolution of the simulation domain.

\subsection{Pore-scale simulation with a porous medium generated by a particle simulation}
To construct a porous structure, we use the in-house multi-body dynamics framework PE \citep{Preclik:2015}. The PE can simulate the motion of rigid bodies and their interaction by frictional collisions. Here we use this functionality to generate a random sphere packing by letting random spheres fall into the simulation domain from the top. After the spheres have come to a rest, their position is fixed and their geometry defines the solid matrix of a porous structure. The pore space is then resolved by a lattice Boltzmann grid.

The particles have possibly different radii that vary up to 50\% of a mean diameter and for each sphere it is chosen randomly. For the fluid flow simulation using the LBM, the TRT collision operator and the CLI solid boundary condition are used. This combination is fast, has second order accuracy and shows no viscosity-dependency.

\begin{figure}
\centering
\includegraphics[trim= 5mm 5mm 12mm 15mm,clip,width=0.5\textwidth]{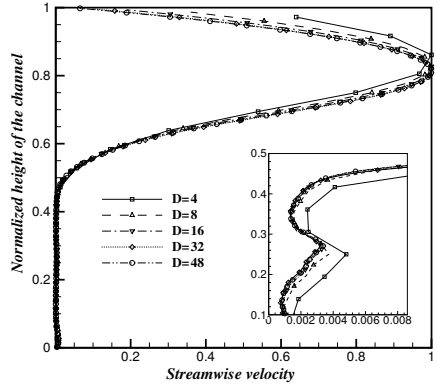}
  \caption{Planar average stream-wise velocity for different grid sizes, $Re_D \simeq 2$.
   }
  \label{fig:DNSGridStudy}
\end{figure}

First, we test the influence of the cell size on the averaged stream-wise velocity. To do so we increase the diameter $D$ of the spheres from $4$ to $48$ and keep $Re_D = \frac{U_{max} D}{\nu}$ constant. The domain has two walls at the top and bottom, and periodic boundary conditions are applied at stream-wise and span-wise directions. A constant pressure drop drives the flow, and the data are set such that $Re_D \simeq 2$. 
The simulation result is presented as a planar average of stream-wise velocity in Fig. \ref{fig:DNSGridStudy} while it is normalized based on the maximum velocity and the height of the channel. As it can be seen, the results converge and for $D=32$ lattice units,
and a further increase of the resolution does not significantly change the results. It is worth to note that in the porous region a coarse lattice can be used and that only the transient region requires a better resolution.

\begin{figure}
\centering
    \subfigure[pore geometry]{
      \label{fig:Monosized}
      \includegraphics[trim= 125mm 5mm 125mm 3mm,clip,width=0.42\textwidth]{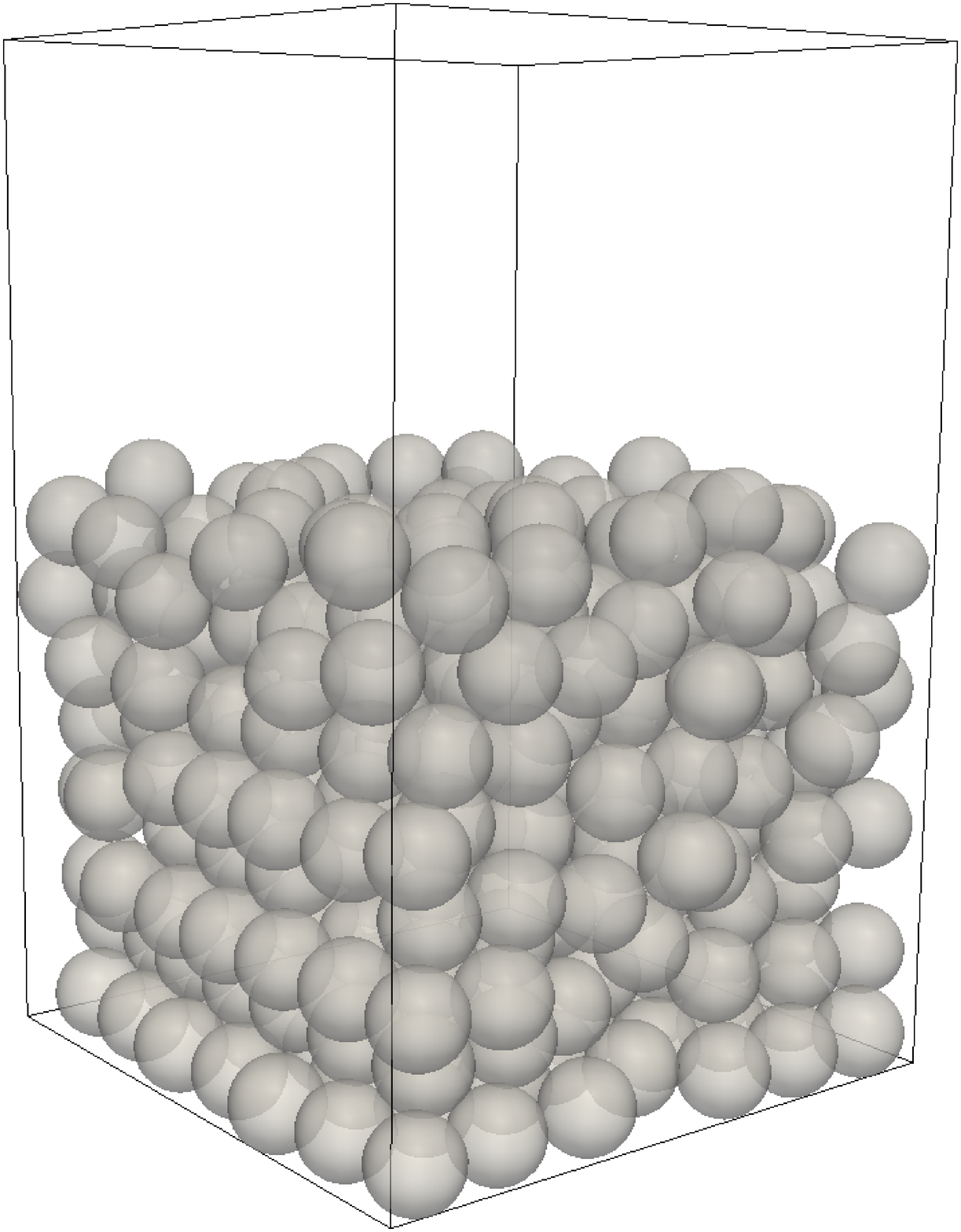}
    }
        \subfigure[planar average of stream-wise velocity]{
      \label{fig:DNSReEffect}
      \includegraphics[trim= 5mm 5mm 12mm 5mm,clip,width=0.5\textwidth]{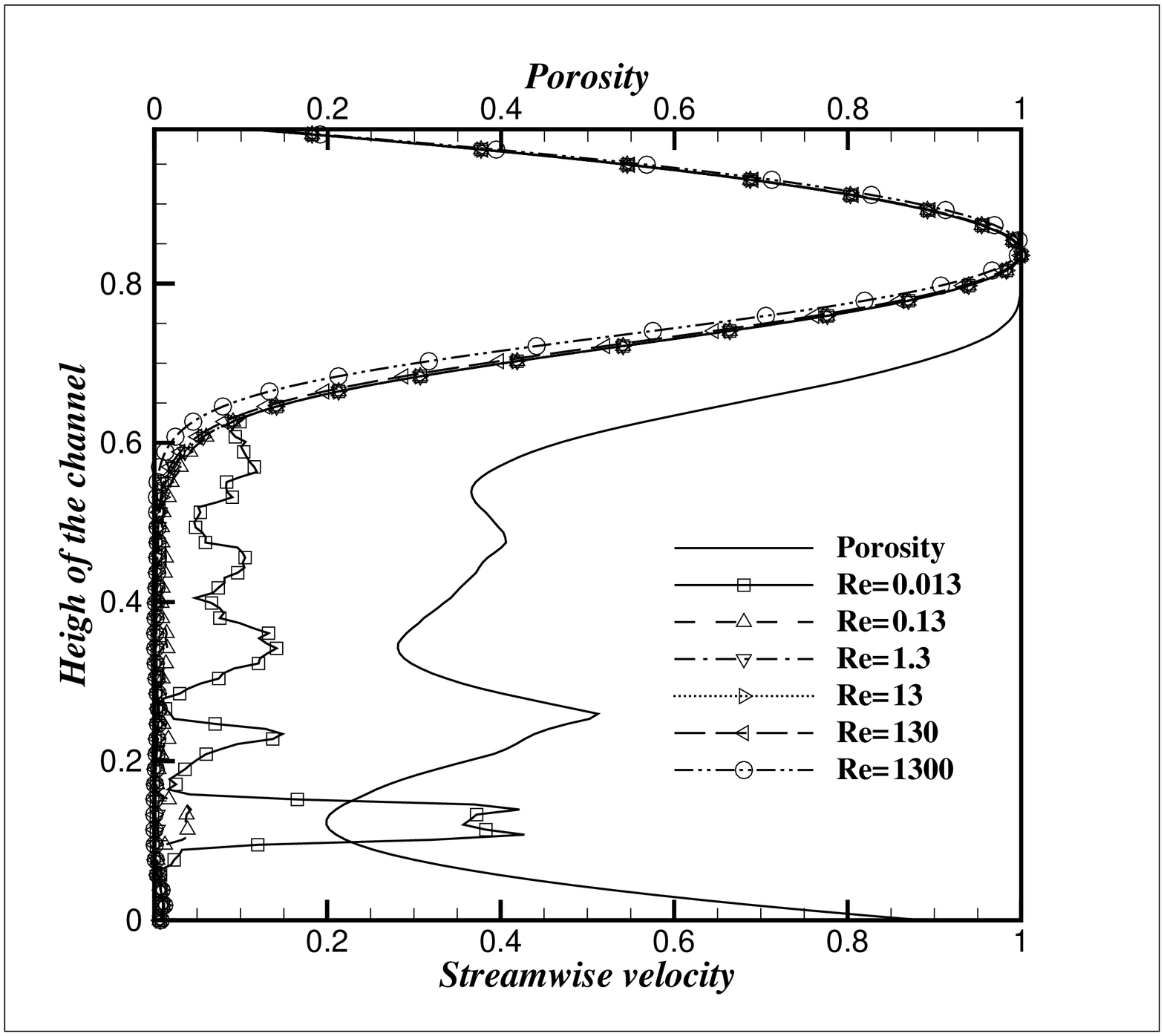}
    }
  \caption{Flow over mono-sized particles for different $Re$ numbers.
   }
  \label{fig:DNSReEffects}
\end{figure}

Fig. \ref{fig:DNSReEffect} shows the planar average stream-wise velocity for different $Re$ numbers. To change the $Re$ number, the viscosity and particles diameter are kept constant while the pressure gradient is changed to adjust the flow velocity. The results show that for slow flow, the velocity in the porous region is considerably higher than for fast flow. When the $Re$ number of the flow increases, the position of the maximum velocity shifts toward to the top wall. This observation results from a boundary layer effect; when the flow velocity is high in the free flow, the penetration to the porous region is less, therefore, the position of the maximum velocity changes.

In Fig. \ref{fig:DNSReEffect}, we observe a small deviation in the velocity profile close to the bottom wall in the porous region. This is because of the high porosity close to the wall, where spherical particles are on a flat plane, see Fig. \ref{fig:Monosized}. Consequently a higher permeability can be found in this region, and the flow will accelerate because the resistance against the pressure difference is lower than in the interior of the porous medium. Therefore, to evaluate the existing models without this effect and having a more uniform porosity in the porous region, a different set-up structure is chosen. The bottom plate of the particle simulation is placed about one particle size below the bottom wall of the fluid flow simulation. With this structure the porosity does not have the effect of placing a sphere on the wall, and therefore we have approximately a uniform permeability distribution in the porous media. 

\begin{figure}
\centering
    \subfigure[pore geometry and streamlines]{
      \label{fig:Str_2}
      \includegraphics[trim= 125mm 5mm 125mm 43mm,clip,width=0.44\textwidth]{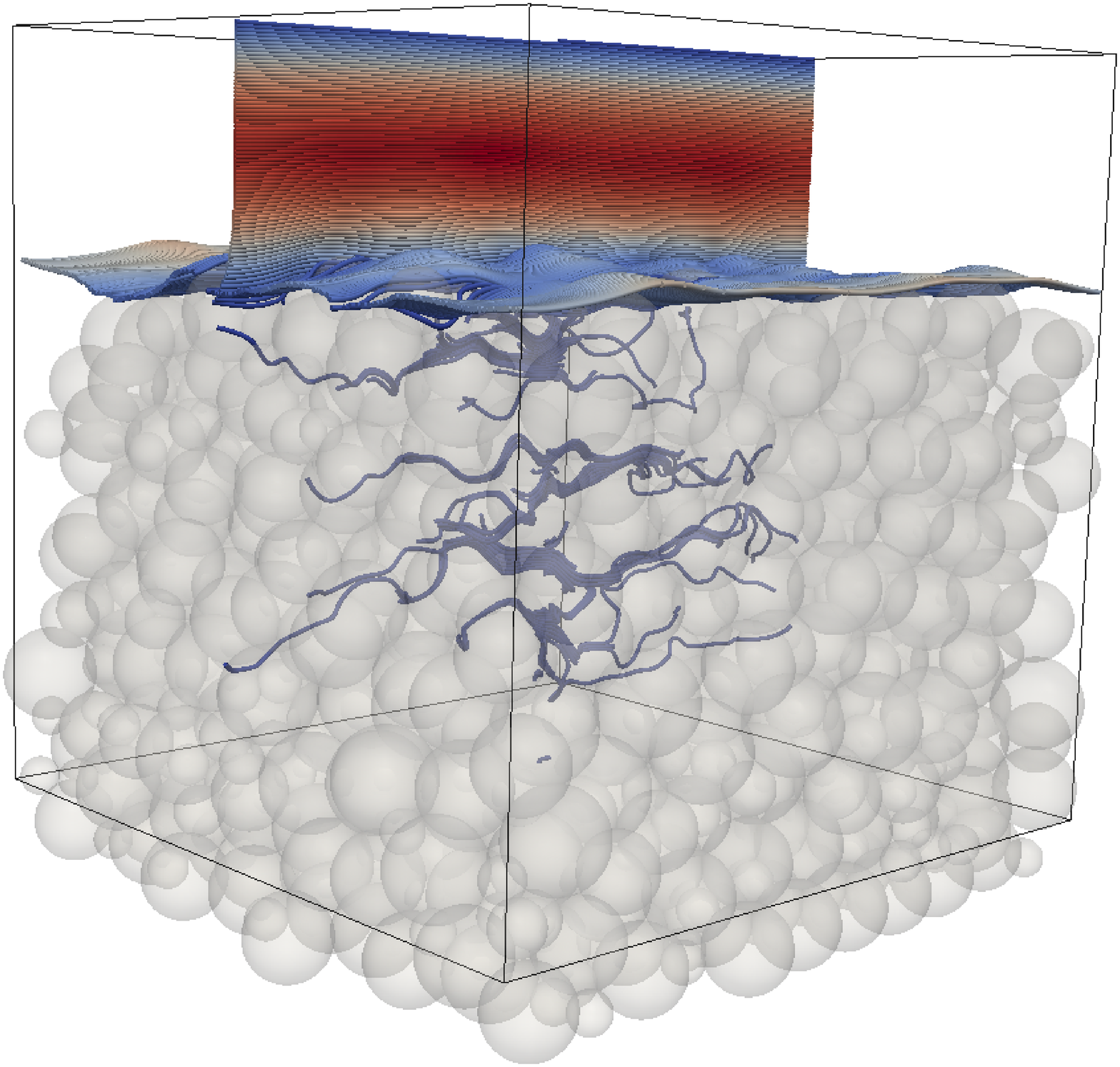}
    }
        \subfigure[planar average of stream-wise velocity]{
      \label{fig:DNSProfile}
      \includegraphics[trim= 5mm 5mm 12mm 5mm,clip,width=0.48\textwidth]{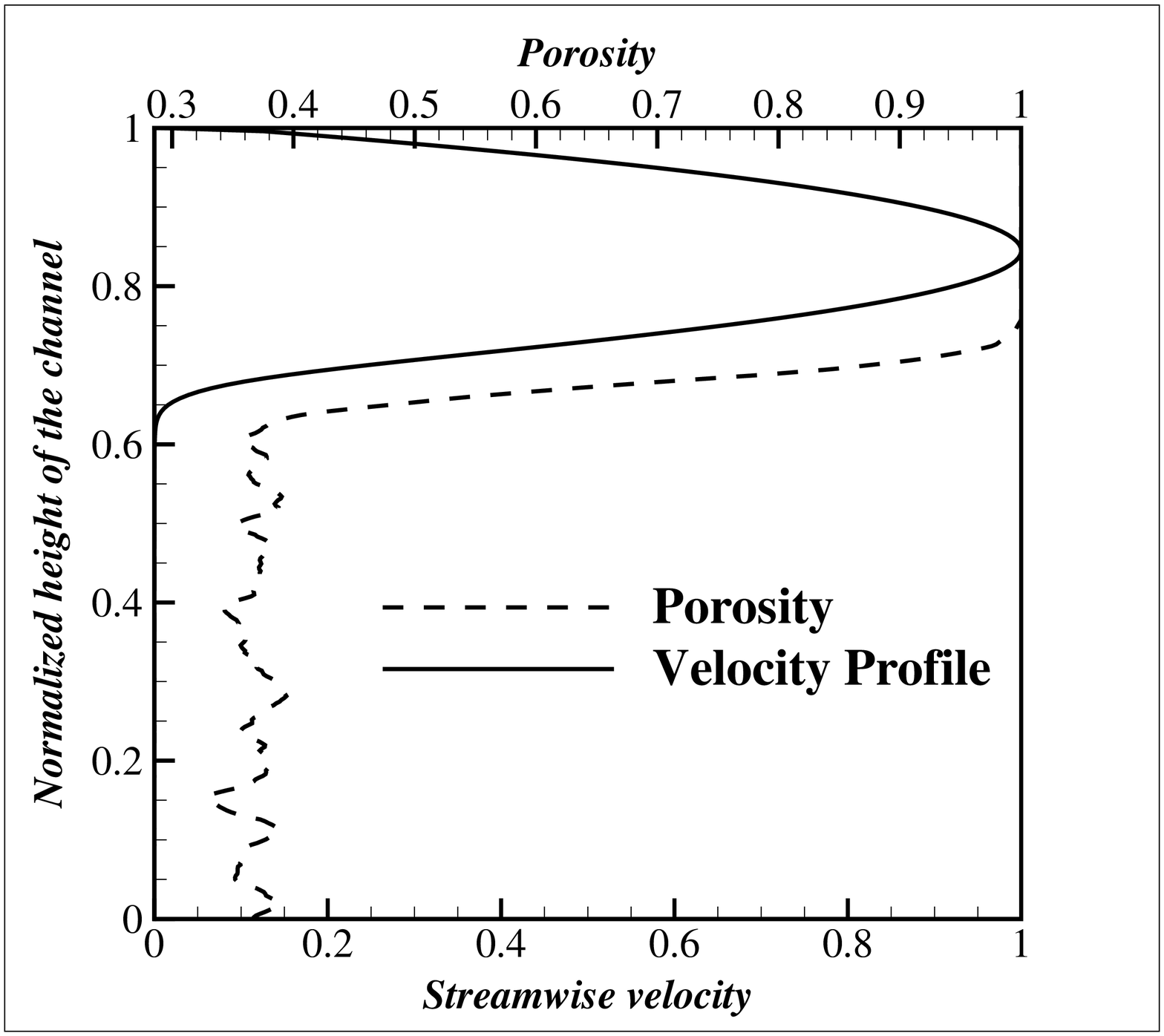}
    }
  \caption{pore-scale simulation of free flow over porous media.
   }
  \label{fig:DNSResult}
\end{figure}

The results of this pore-scale simulation are taken as a reference solution. Here, we use 1274 particles with diameters in the range of 16-48 cells. The flow is driven by a pressure difference of $10^{-6}$ (in lattice units), and the simulation is run until the flow reaches the steady state. The planar average of the stream-wise velocity is depicted in Fig. \ref{fig:DNSResult}.
\subsection{Evaluation of different interface conditions }
In this subsection, we evaluate different two-domain approaches.
All interface conditions under consideration have parameters for which no explicit relation is known. In the BJ and BJS models, the slip coefficient, $\alpha$, is unknown, while in the OTW model, the jump coefficient $\beta$ and the effective viscosity $\mu_{\rm eff}$ are unknown and in the Br model, the effective viscosity $\mu_{\rm eff}$ is unknown. 

\begin{figure}[h]
\centering
\includegraphics[width=0.85\textwidth]{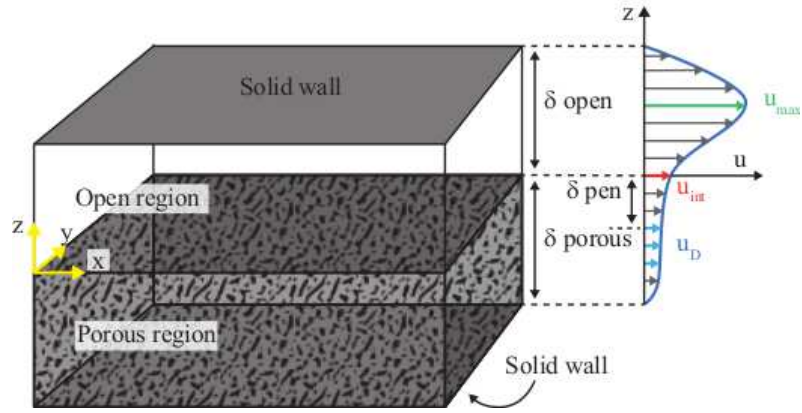}
  \caption{Schematic of the simulation domain and averaged velocity profile in the open and porous regions.
   }
  \label{fig:DomainPorous}
\end{figure}

By using the DNS solution, we calculate the optimal value for the unknown parameters.
The domain that is used is a channel which is periodic in stream-wise and span-wise directions (Fig. \ref{fig:DomainPorous}). A free fluid flows on the top of a porous media. To have a good comparison, all of the flow properties are non-dimensionalized. 

The value of the interface velocity $U_{int}$, can be directly obtained from the averaged velocity profile of the DNS. In order to obtain the velocity gradient on the open and porous sides, curve fitting techniques are used to approximate the velocity profile close to the interface. The velocity profile on the open side can be well approximated by a polynomial curve and on the porous side, the velocity profile can be approximated by an exponential curve. Permeability and seepage velocity (Darcy velocity) can be calculated from the velocity profile far from the interface in the porous medium. Given this, the unknown variables can be calculated from the Eqs. \eqref{eq:BJ}, \eqref{eq:BJS} and \eqref{eq:WTK}. However, to do so, the exact position of the interface should be defined which in real applications is nearly impossible.

\begin{figure}
\centering
    \subfigure[Br]{
      \label{fig:Analy_Brink}
      \includegraphics[trim= 5mm 3mm 10mm 10mm,clip,width=0.4\textwidth]{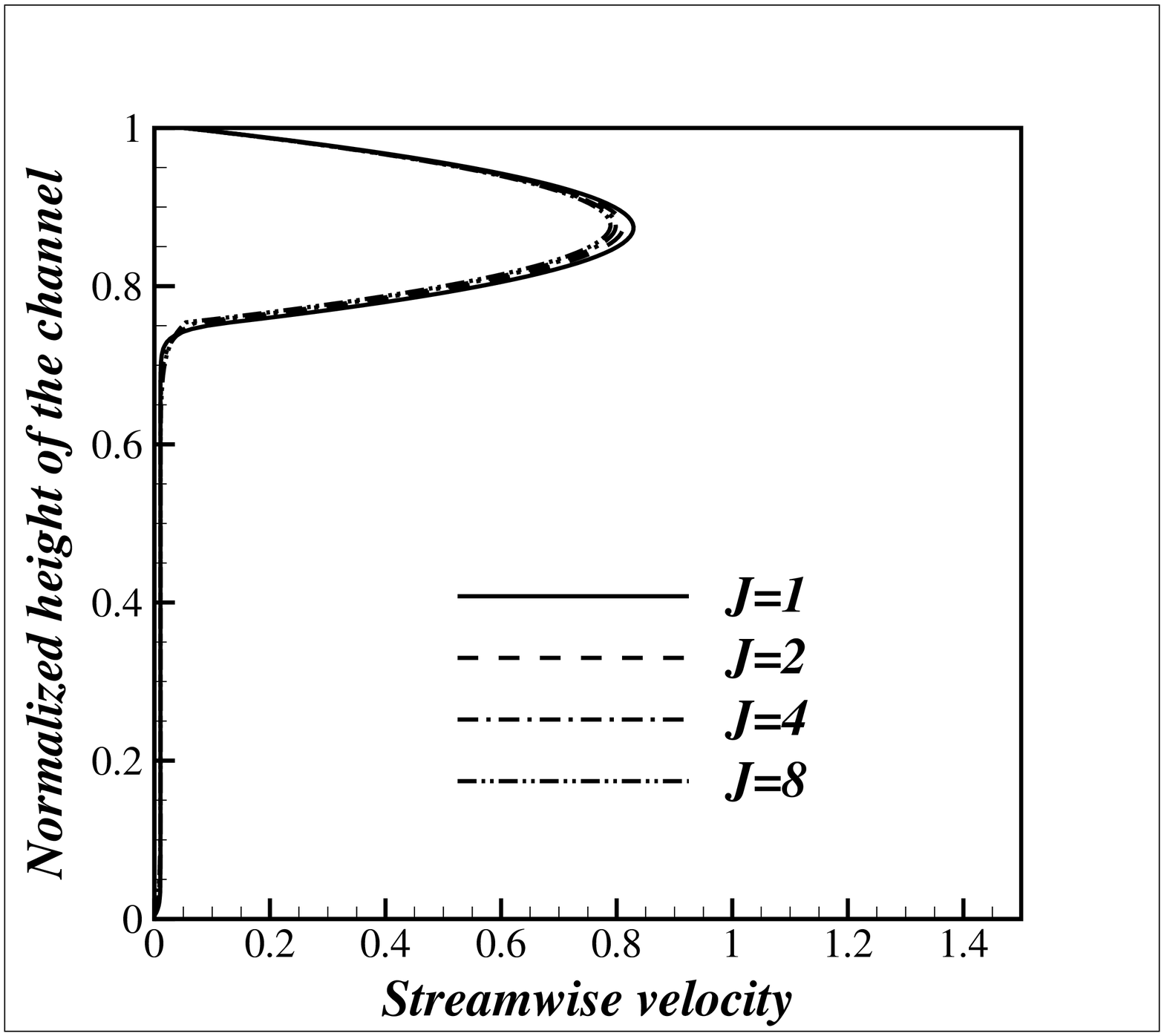}
    }
    \subfigure[OTW]{
      \label{fig:Analy_OTW}
      \includegraphics[trim= 5mm 3mm 10mm 10mm,clip,width=0.4\textwidth]{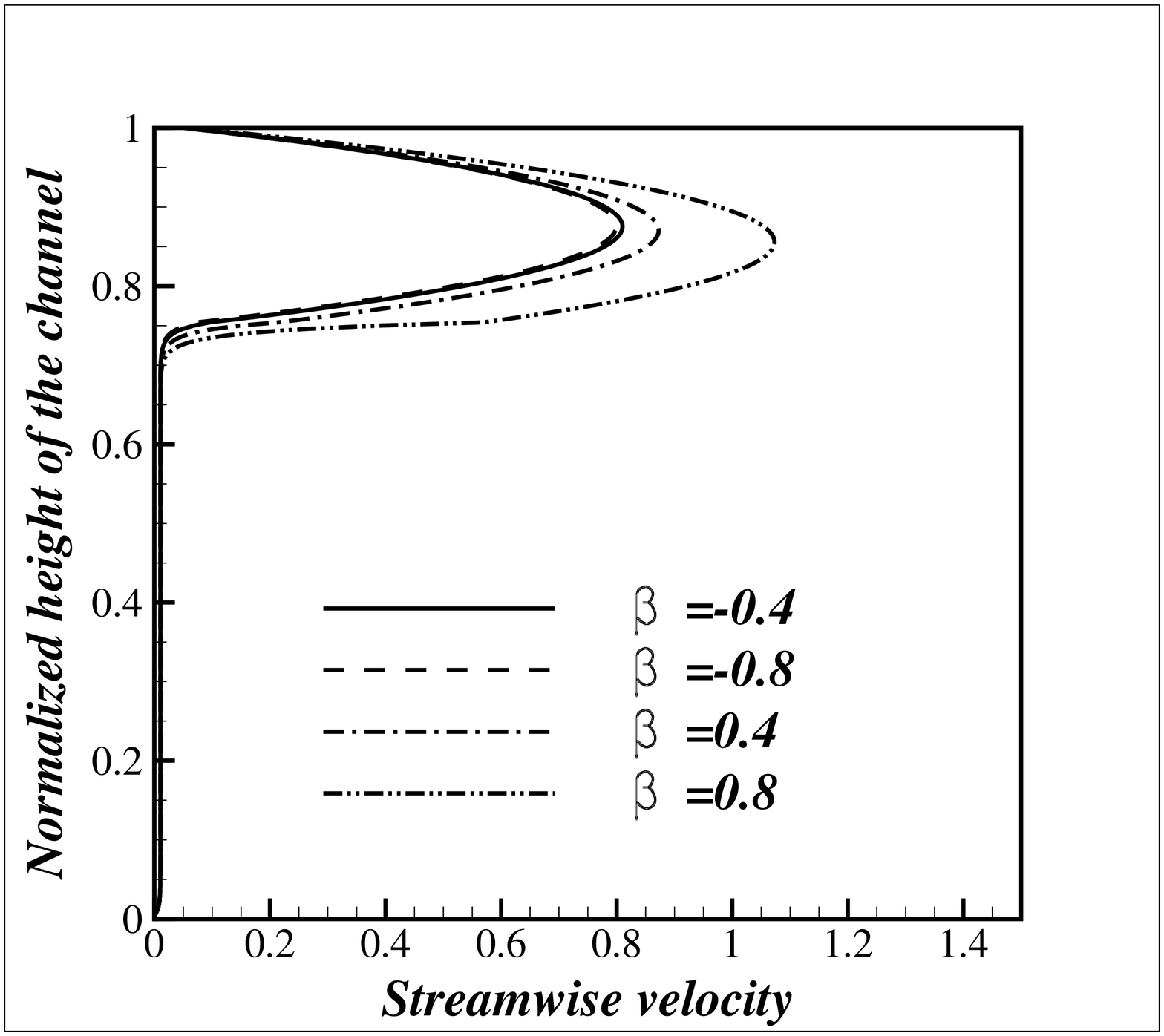}
    }\\
        \subfigure[BJ]{
      \label{fig:Analy_BJ}
      \includegraphics[trim= 5mm 3mm 10mm 10mm,clip,width=0.4\textwidth]{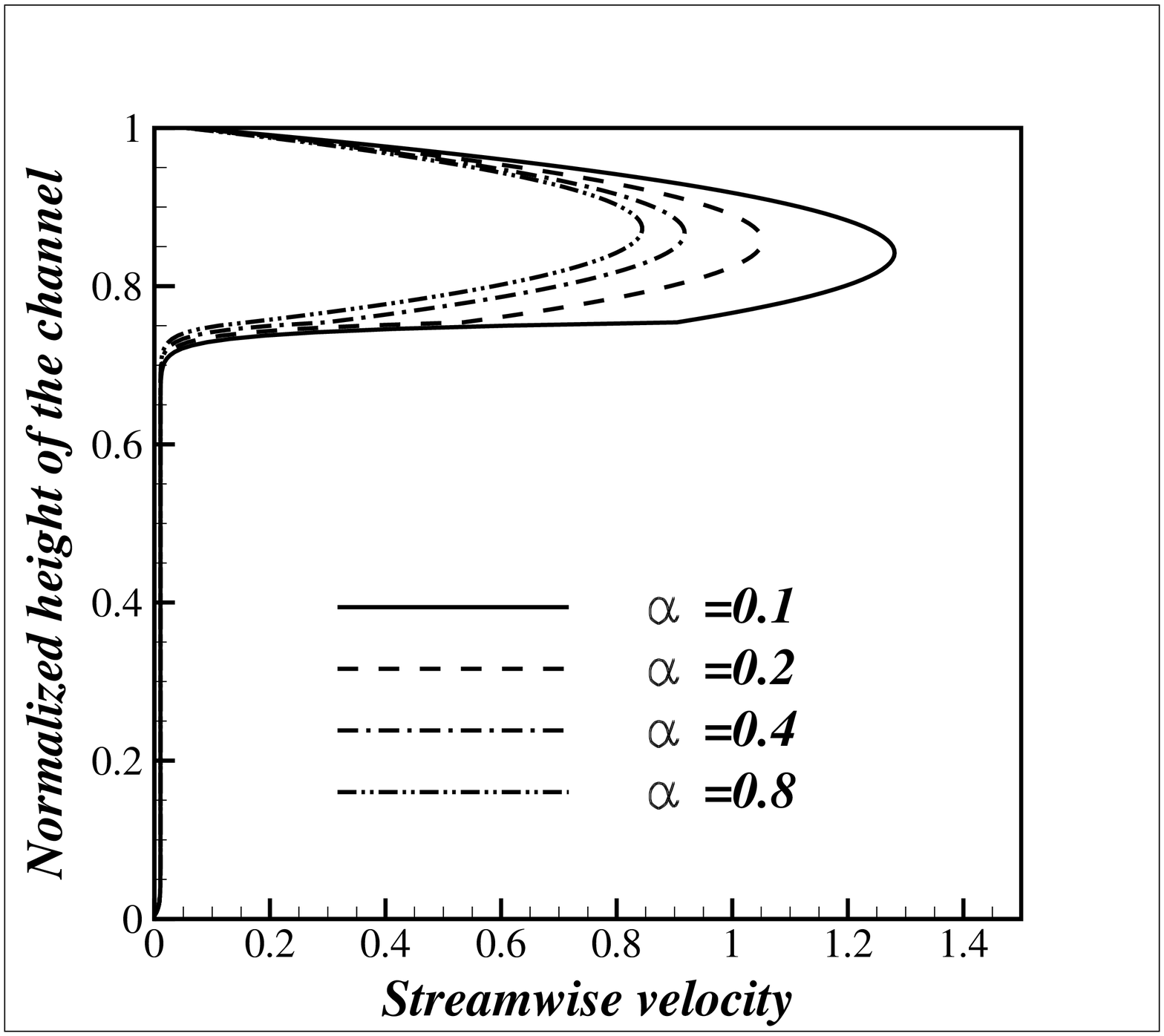}
    }
    \subfigure[BJS]{
      \label{fig:Analy_BJS}
      \includegraphics[trim= 5mm 3mm 10mm 10mm,clip,width=0.4\textwidth]{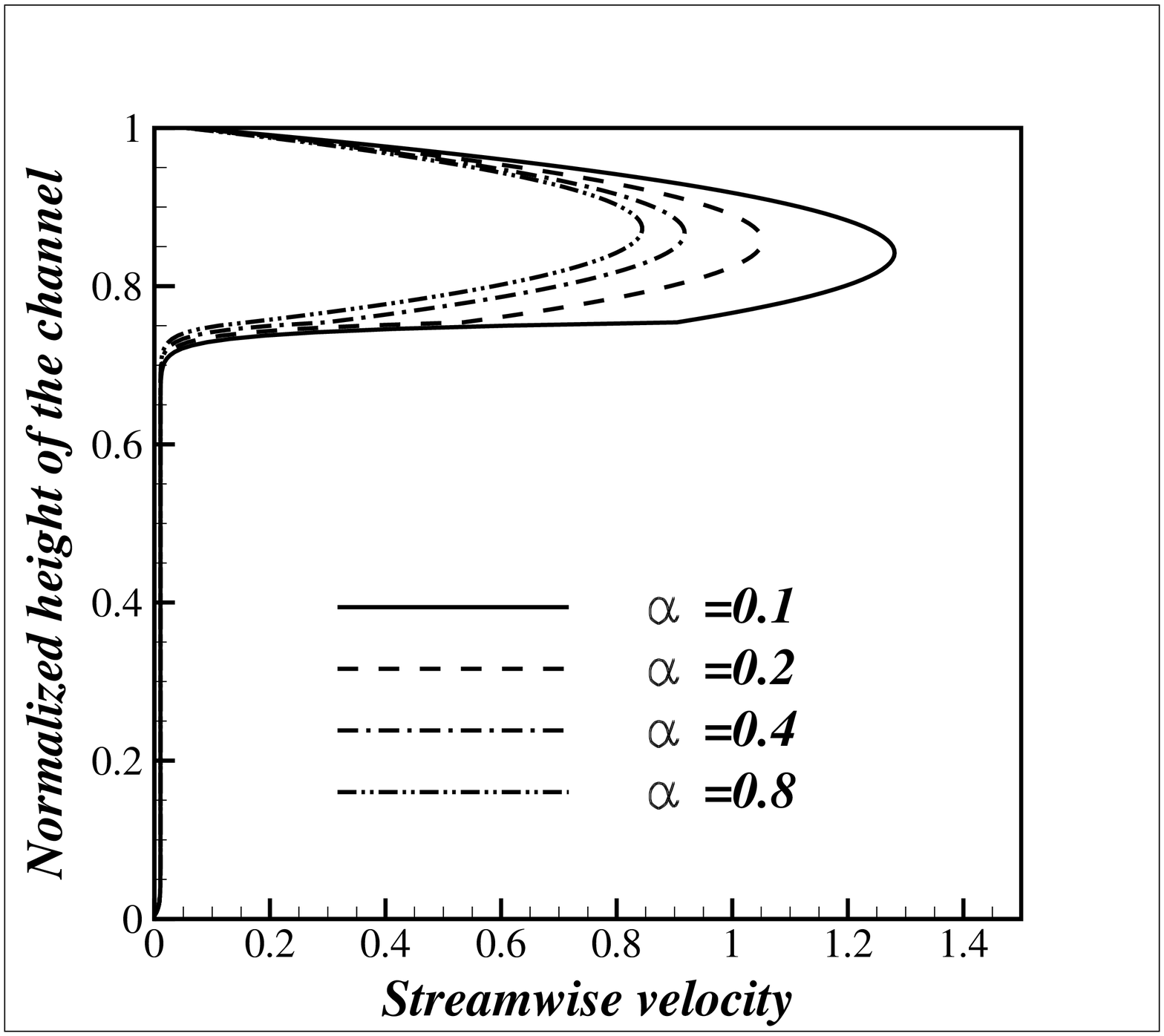}
    }
  \caption{Analytical solution for the velocity profile, which is normalized by the maximum velocity of the DNS solution, by different interface models.
   }
  \label{fig:Analytical_param}
\end{figure}

To find out how the additional parameters of the interface conditions
 affect the results, a two-domain approach is chosen and solved analytically. For the free flow region, the Stokes equation is used and for the porous region, the Brinkman's equation is chosen. The permeability is calculated from the DNS result far enough from the interface inside the porous region. In Fig. \ref{fig:Analytical_param}, we depict the planar average stream-wise velocity which is normalized based on the maximum velocity in the DNS solution.

As it can be seen in Fig. \ref{fig:Analy_Brink}, in the Brinkman model by increasing the viscosity ratio, $J=\frac{\mu_{\rm eff}}{\mu}$, the maximum velocity decreases and produces a discontinuity in the shear stress over the interface. In the OTW model (Fig. \ref{fig:Analy_OTW}), negative values of $\beta$ do not influence the result significantly, however, positive value of $\beta$ have a strong impact on the maximum velocity as well as on the slip velocity on the interface. Fig. \ref{fig:Analy_BJ} and Fig. \ref{fig:Analy_BJS} show the results for the BJ and the BJS interface conditions. It can be observed that there is almost no difference between these two models for low $Re$ number flows. In both these cases, the maximum velocity decreases if $\alpha $ increases. A small value of $\alpha$ results in a considerably larger maximal velocity than in the two other cases.

\begin{figure}
\centering
    \subfigure[Interface at y=0.756]{
      \label{fig:172}
      \includegraphics[trim= 5mm 3mm 10mm 10mm,clip,width=0.4\textwidth]{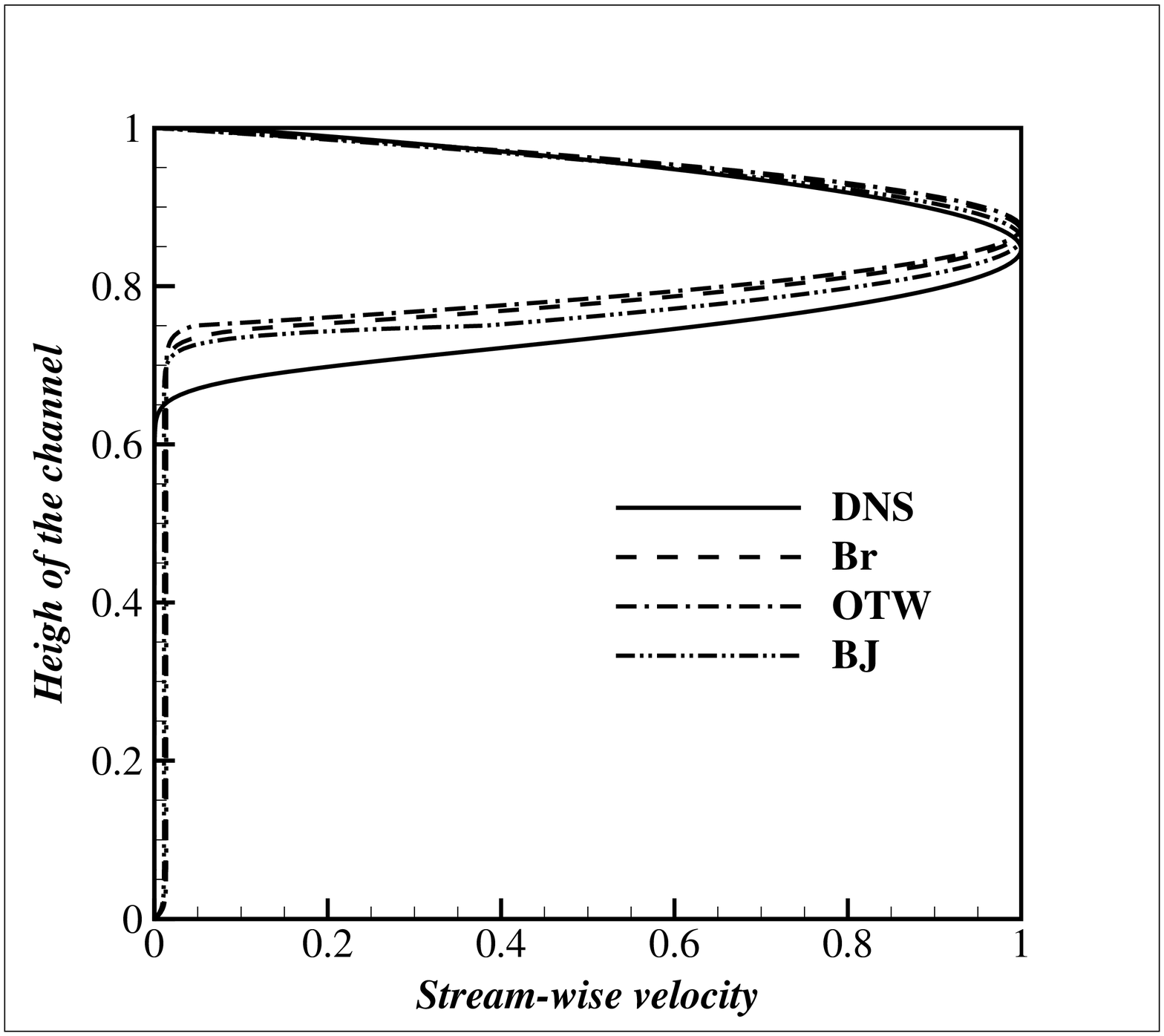}
    }
    \subfigure[Interface at y=0.722]{
      \label{fig:180}
      \includegraphics[trim= 5mm 3mm 10mm 10mm,clip,width=0.4\textwidth]{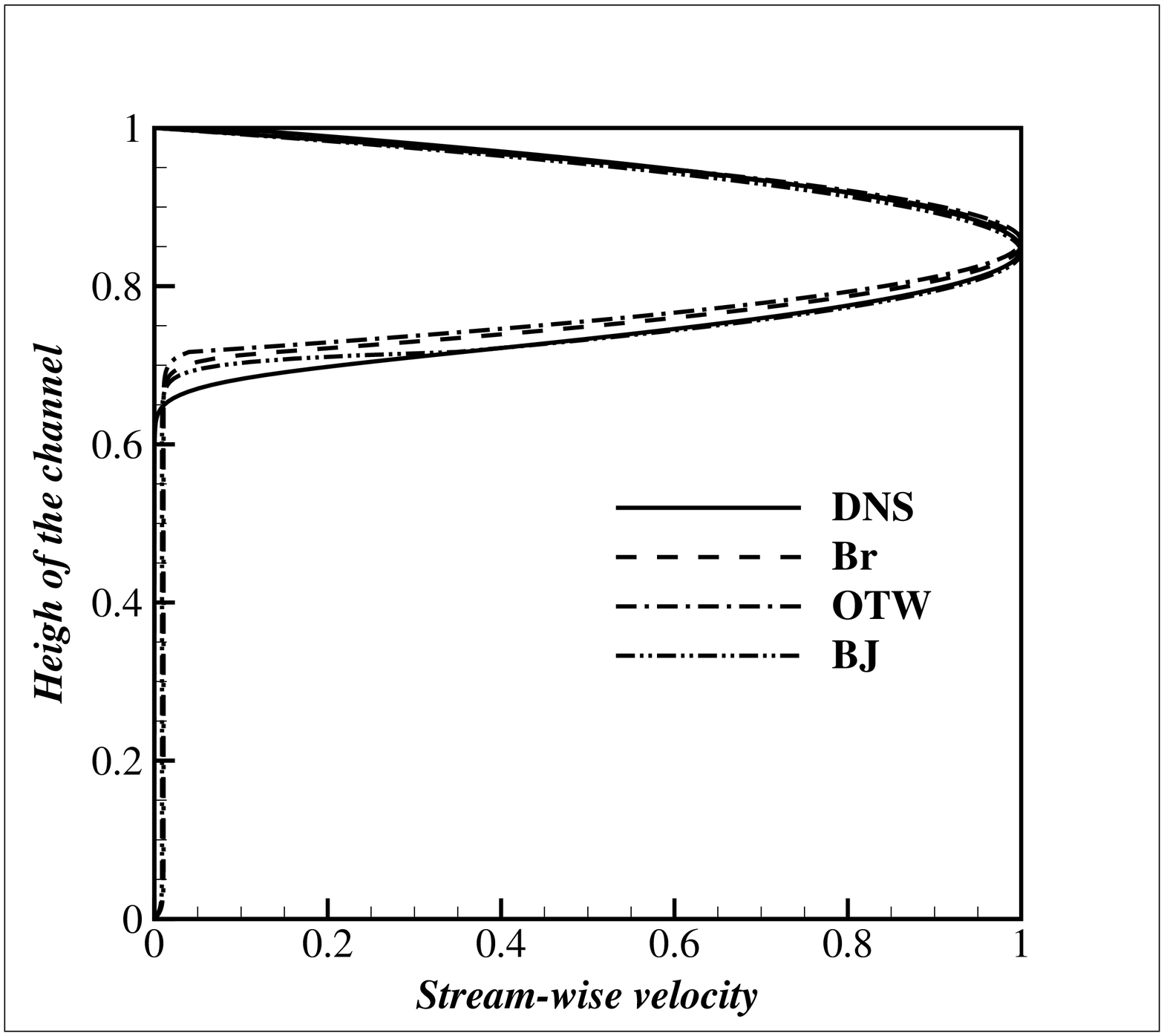}
    }
  \caption{Normalized velocity profile of the one-domain approaches in compare to the DNS solution; a) interface at y=0.756, b) interface at y=0.722.
   }
  \label{fig:SharpvsDNS}
\end{figure}

Quite often two-domain models result in discontinuities in the stress 
at the interface. Thus the a priori knowledge of the position of the interface is crucial. One possibility to fix the position of the interface
is to take the location where the porosity reaches the limit value one, i.e., $y=0.756$. However fitting of the DNS velocity profile shows that only up to $y=0.722$, the curve is fitted well by an exponential function. More precisely, $y = 0.48423 \cdot \exp(0.31195\, x) - 0.48236 \cdot \exp(0.3131\, x)$ yields a root mean squared error of $5.736 \cdot 10^{-6}$. The pure fluid flow velocity profile is fitted to a 2nd order polynomial resulting in $y = (1.9593 e-3) + (2.78421 e-4)x - (4.48066 e-6)x^2$ with a root mean squared error of $9.5815 \cdot 10^{-6}$. This observation motivates an alternative choice of the interface position. Calculating the slip coefficient and the jump coefficient for these two positions, we find for $y=0.756$, $\alpha=0.3163$, $\beta=-2.8397$ and for $y=0.722$, $\alpha=0.31645$ and $\beta=-2.8397$. However, as it can be seen in Fig. \ref{fig:SharpvsDNS}, even with the parameters which are extracted from the DNS results, the considered two-domain approaches cannot represent accurately the DNS solution. Comparing Figs. \ref{fig:172} and \ref{fig:180} shows that the two-domain approaches depend strongly on the interface position and more sophisticated 
criteria for defining the interface location are required to obtain better matching results.

\section{Comparsion of a homogenized LBM with the pore-scale LB  simulation}
Different models for isothermal incompressible fluid flow in porous media are proposed by several groups. In this work, we use the generalized lattice Boltzmann model (GLBM) for porous media introduced in \citep{Guo:2002}, which is applicable for a medium with both a constant and a variable porosity. The model can be expressed by the following generalized Navier-Stokes equation: 
\begin{equation}
	\label{eq:GeneralizedPor1}
	\nabla \cdot \mathbf{u}=0
\end{equation}
\begin{equation}
	\label{eq:GeneralizedPor2}
	\frac{\partial \mathbf{u}}{\partial t} + \left(\mathbf{u} \cdot \nabla \right) \left(\frac {\mathbf{u}} 	{\epsilon}\right) = - \frac{1}{\rho} \nabla \left(\epsilon p \right) + \nu_{\rm eff} \nabla^2 \mathbf{u} + 	\mathbf{F},
\end{equation}
where $\rho$ is the fluid density, $ \textbf{u}$ and $p$ are the volume-averaged velocity and pressure, respectively, $\nu_{\rm eff}$ is the effective viscosity, and $ \epsilon $ is the porosity. The total body force $\textbf F$ caused by the presence of a porous medium and other external force fields is given by

\begin{equation}
	\label{eq:PorousForce}
	\mathbf{F}= - \frac{\epsilon \nu}{K}\mathbf{u} - \frac{\epsilon c_F }{\sqrt{K}} | \mathbf{u} | \mathbf{u} + \epsilon \mathbf{G},
\end{equation}
where $\nu$ is the shear viscosity of the fluid that is not necessarily the same as $\nu_{\rm eff}$, $\textbf G$ is the body force induced by an external force, $c_F$ is the Forchheimer coefficient that depends on the porous structure, and $K$ is permeability of the porous media. The first and the second terms on the right hand side of Eq. \eqref{eq:PorousForce} are the linear Darcy and non-linear Forchheimer drags due to the porous medium, respectively. The quadratic nature of the non-linear resistance makes it negligible for low-speed flows, but is more noteworthy in hindering the fluid motion for high-speed flows, i.e., high $Re$ number  and high $Da$ number flows. 

Firstly to validate the generalized model for flow over a porous
medium, we choose a simple Couette flow.
The lower-half of the channel of width $H$ is filled with a porous medium with a porosity of $ \epsilon$, the stream-wise and span-wise boundaries are periodic, and the top wall of the channel is moving with a constant velocity of $u_0$.
Then, the steady state velocity in this channel satisfies the equation 

\begin{equation}
	\label{eq:SteadyVelocity}
	\nu_{\rm eff} \nabla^2 \mathbf{u} - \frac{\epsilon \nu}{K}\mathbf{u} - \frac{\epsilon c_F }{\sqrt{K}} | \mathbf{u} | 	\mathbf{u} + \epsilon \mathbf{G} = 0,
\end{equation}
while the walls of the channel are modeled by a no-slip condition.

\begin{figure}[ht]
\centering
        \subfigure[]{
            \label{fig:JEffectfull}
            \includegraphics[trim= 2mm 5mm 10mm 10mm,clip,width=0.4\textwidth]{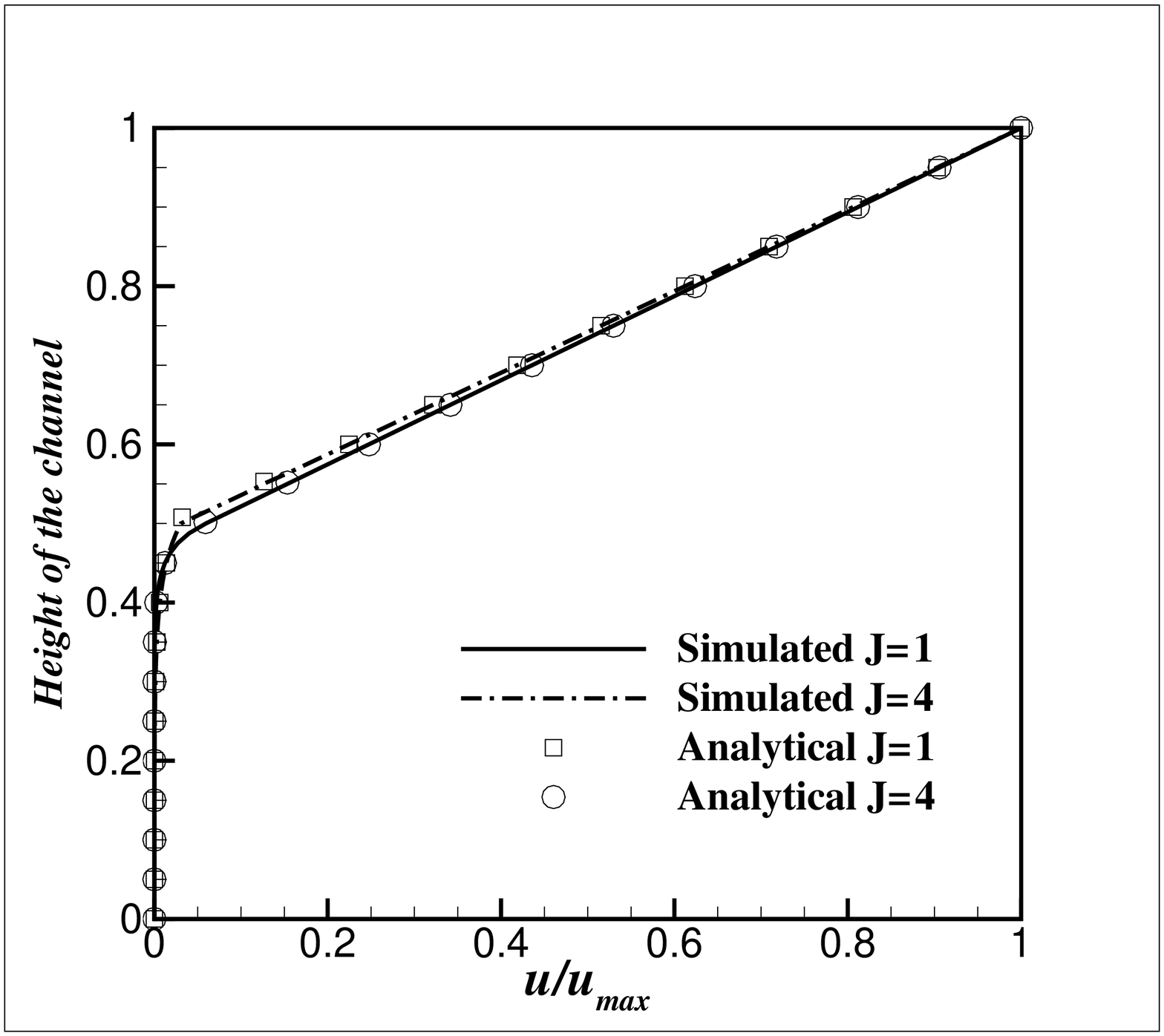}
                    }
        \subfigure[]{
           \label{fig:JEffectPart}
           \includegraphics[trim= 2mm 5mm 10mm 10mm,clip,width=0.4\textwidth]{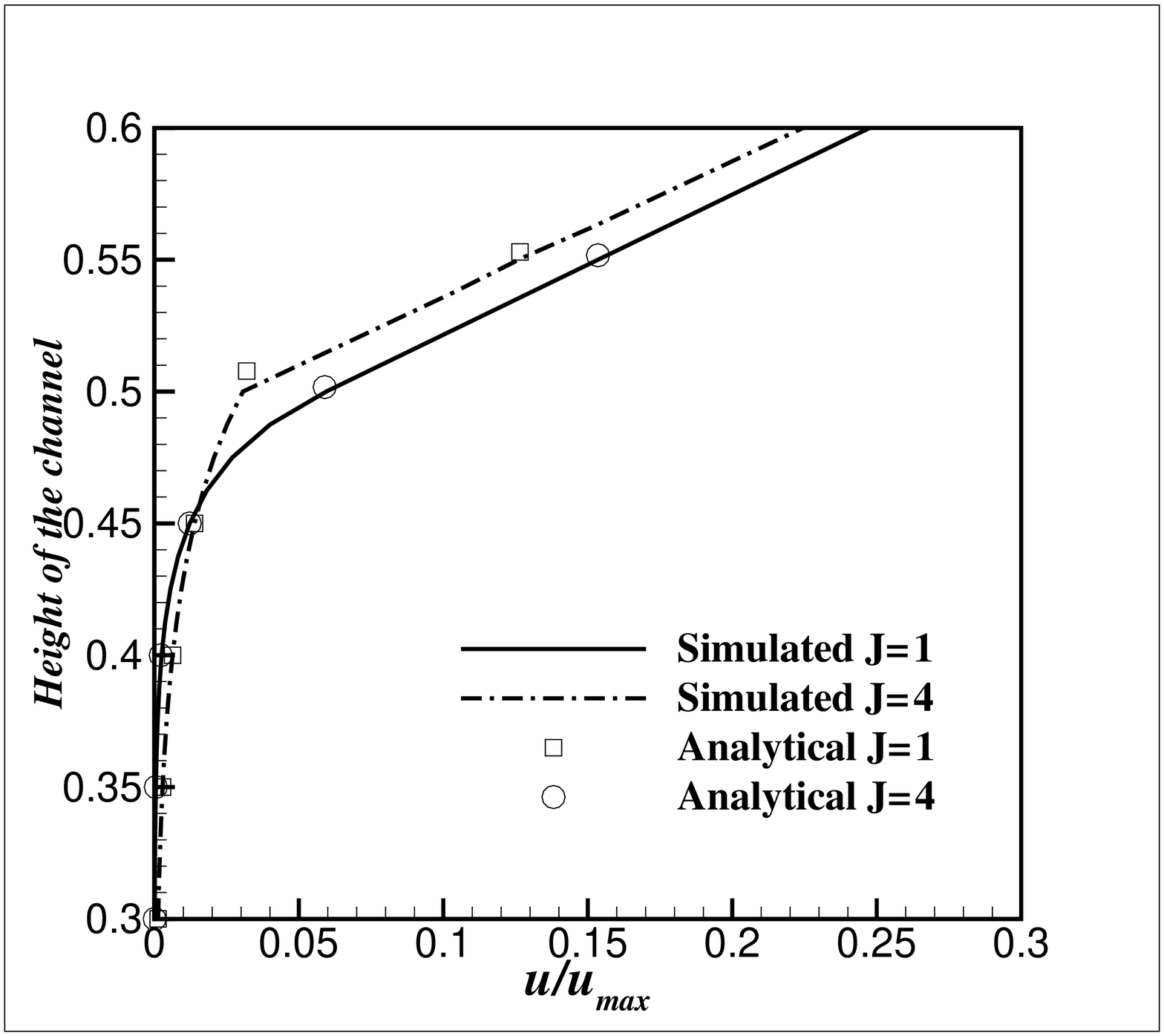}
        }
    \caption{Velocity profile of the Couette flow for different viscosity ratios $J=\mu_e/\mu$, in comparison with the approximate analytical solution of Eq. \eqref{eq:Qouette}, a) global system; b) zoom into the region near the interface
     }
   \label{fig:QouetteSoloution}
\end{figure}

Fig. \ref{fig:QouetteSoloution} shows the velocity profile for the Couette flow with different viscosity ratios J ($=\mu_e/\mu$) and compared to a semi-analytical solution for $Re=0.1$ and $Da=0.00012$.
 In the Stokes regime for low $Da$ number, \citep{Martys:1994}
 reported that the velocity profile in the free flow is linear and
 exponentially decaying in the porous region. More precisely the
 semi-analytic solution can be written as:
\begin{equation}
	\label{eq:Qouette}
	u_x(y) = \begin{cases} rKa + \epsilon a \left(y - H/2 \right) &\quad {H/2 \leq y \leq H}\\
	rKa e^{r(y-H/2)} & \quad {0 \leq y \leq H/2} \end{cases}
\end{equation}
while
\begin{equation}
	a= \frac{2 u_0}{2rK+\epsilon H}, \quad r= \frac{\sqrt{ \nu \epsilon}}{\sqrt{  \nu_{\rm eff} k}},
\end{equation}
and  $ u_0$ is the lid's velocity. 
 The simulation result shows excellent agreement with the analytical solution for both viscosity ratios.

Secondly, we apply the generalized model to  a problem with no sharp
interface and a significant  porosity change close to the interface. We use the planar average of the porosity as it is obtained in the DNS, therefore, there is no need to explicitly set the interface position. Since the flow is within the Stokes regime, the Forchheimer term in Eq. \eqref{eq:PorousForce} is neglected.

\begin{figure}[h!]
\centering
\includegraphics[trim= 5mm 5mm 12mm 5mm,clip,width=0.5\textwidth]{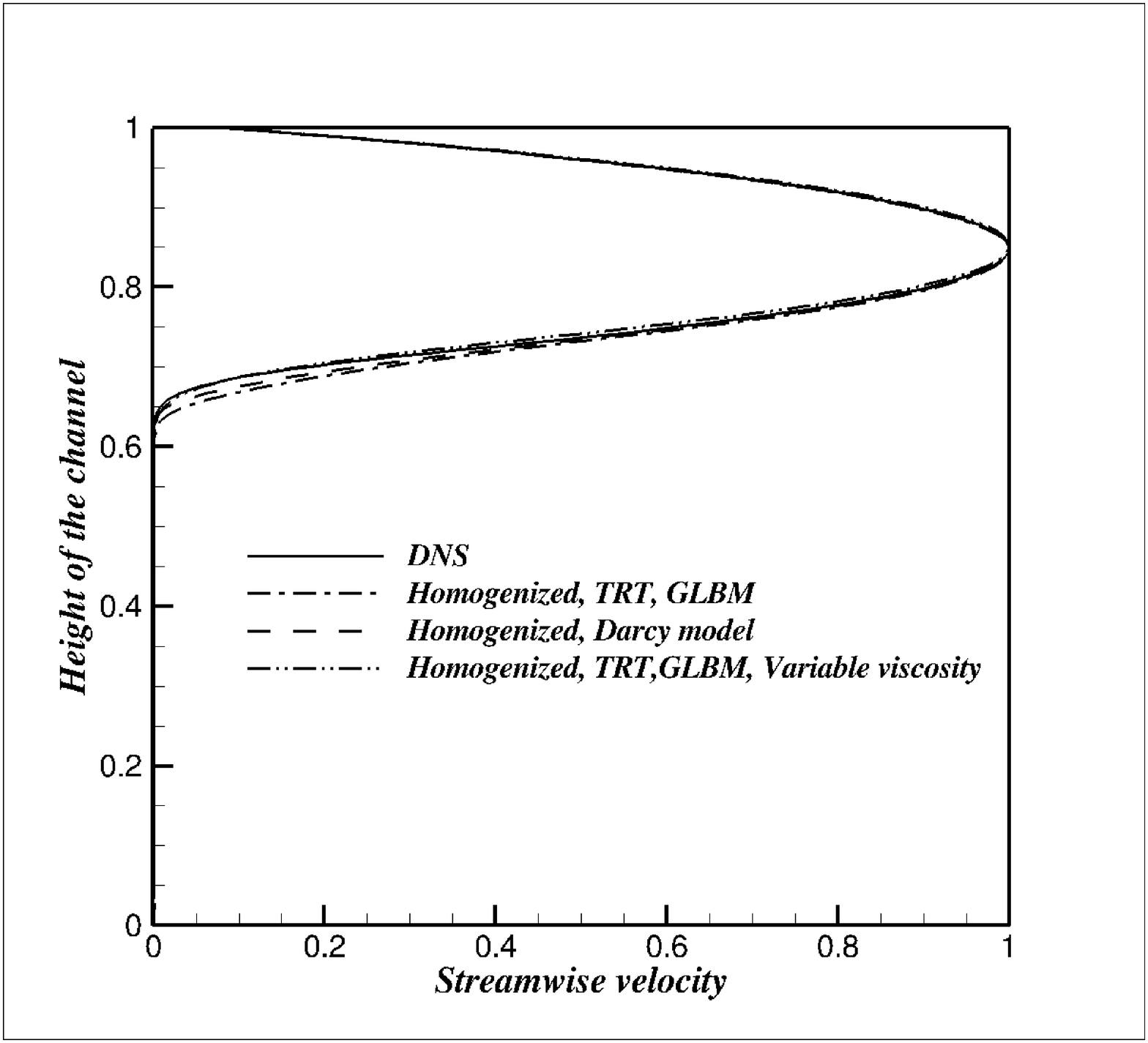}
    \caption{A comparison between the planar average of the stream-wise velocity obtained by DNS and the homogenized model, $Re_D \simeq 2$.
     }
    \label{fig:DNSvsHomo}
\end{figure}

Fig. \ref{fig:DNSvsHomo} shows the results of the planar average stream-wise velocity for the DNS solution and the GLBM. Although the porosity, permeability, fluid properties and driving forces are the same, the standard GLBM homogenized model over-predicts the velocity in the transition zone. The dashed line shows the homogenized model that only takes the Darcy force into account. These two mentioned homogenized models use a viscosity in the porous region which is equal to the free flow region. We propose to use the GLBM homogenized model but  with a viscosity in the porous region depending on the  porosity by $\mu_{\rm eff} = {\mu}/{\epsilon}$. As we can observe in the porous region, the latter model  can perfectly predict the DNS result.

\section{Conclusion}
We presented three different approaches to simulate the interaction of free flow with porous media flow, namely, direct pore-scale simulations, as well as homogenized single-domain and two-domains approaches. The lattice Boltzmann method is employed both, for obtaining the pore-scale reference solution, and for solving the computationally more appealing homogenized problems.

For the two-domain approaches, four different interface conditions for dealing with the physical transport through a sharp interface have been evaluated.
Our comparison yields that the two-domain techniques are quite sensitive to the interface position. To further investigate this effect, we examined two definitions for the interface position, i.e., the exact and the apparent position assumptions. However, as our results indicate, both approaches fall short with respect to accuracy in the vicinity of the interface if the exact interface geometry is unknown.
As an alternative approach we consider a homogenized one-domain model that is based on the idea of a smooth transition zone between the free flow and porous media models. A simple porosity-dependent rescaling of the viscosity allows us to accurately reproduce the results obtained by averaging the pore-scale solution.

In future work we aim to investigate the combination of both approaches to allow for the treatment of more general situations in a two-scale fashion. Since the discussed lattice Boltzmann schemes are suitable for REV-scale computations, and are also highly scalable for pore-scale simulations, they lend themselves well for leveraging the power of massively parallel computing architectures.

\section*{Acknowledgement}
\noindent Financial support from the German Research Foundation (DFG, Project WO 671/11-1) and also the International Graduate School of Science and Engineering (IGSSE) of the Technische Universit\"at M\"unchen for research training group 6.03 are gratefully acknowledged. Our special thank goes to Regina Ammer for fruitful discussions and the \waLBerla~primary authors Florian Schornbaum, Christian Godenschwager and Martin Bauer for their essential help with implementing the code.



\begin{thebibliography}{10}

\bibitem{Helmig:2011}
Helmig, R.: Multiphase flow and transport processes in the subsurface: A
  contribution to the modeling of hydrosystems, Springer  (2011)

\bibitem{Alazmi:2001}
Alazmi, B., Vafai, K.: Analysis of fluid flow and heat transfer interfacial
  conditions between a porous medium and a fluid layer, International Journal
  of Heat and Mass Transfer 44, 1735 -- 1749 (2001)

\bibitem{Nield:2009}
Nield, D., Kuznetsov, A.: The effect of a transition layer between a fluid and
  a porous medium: shear flow in a channel, Transport in Porous Media 78,
  477--487 (2009)

\bibitem{Lebars:2006}
Le~Bars, M., Worster, M.G.: Interfacial conditions between a pure fluid and a
  porous medium: implications for binary alloy solidification, Journal of Fluid
  Mechanics 550, 149--173 (2006)

\bibitem{Goyeau:2003}
Goyeau, B., Lhuillier, D., Gobin, D., et~al.: Momentum transport at a
  fluid-porous interface, International Journal Of Heat And Mass Transfer 46,
  4071--4081 (2003)

\bibitem{Chandesris:2009}
Chandesris, M., Jamet, D.: Jump conditions and surface-excess quantities at a
  fluid/porous interface: A multi-scale approach, Transport in Porous Media 78,
  419--438 (2009)

\bibitem{Goharzadeh:2005}
Goharzadeh, A., Khalili, A., Jørgensen, B.B.: Transition layer thickness at a
  fluid-porous interface, Physics of Fluids 17, 057102 (2005)

\bibitem{Ghisalberti:2010}
Ghisalberti, M.: The three-dimensionality of obstructed shear flows,
  Environmental Fluid Mechanics 10, 329--343 (2010)

\bibitem{Morad:2009}
Morad, M., Khalili, A.: Transition layer thickness in a fluid-porous medium of
  multi-sized spherical beads, Experiments in Fluids 46, 323--330 (2009)

\bibitem{Pokrajac:2009}
Pokrajac, D., Manes, C.: Velocity measurements of a free-surface turbulent flow
  penetrating a porous medium composed of uniform-size spheres, Transport in
  Porous Media 78, 367--383 (2009)

\bibitem{Beavers:1967}
Beavers, G.S., Joseph, D.D.: Boundary conditions at a naturally permeable wall,
  Journal of Fluid Mechanics 30, 197--207 (1967)

\bibitem{Nield:2006}
Nield, D., Bejan, A.: Convection in porous media, Springer  (2006)

\bibitem{Duman:2009}
Duman, T., Shavit, U.: An apparent interface location as a tool to solve the
  porous interface flow problem, Transport in Porous Media 78, 509--524 (2009)

\bibitem{Baber:2012}
K.Baber, K.Mosthaf, Flemisch, B., Helmig, R., M\"uthing, S., Wohlmuth, B.:
  Numerical scheme for coupling two-phase compositional porous-media flow and
  one-phase compositional free flow, IMA J. Appl. Math. 6, 887--909 (2012)

\bibitem{Mosthaf:2011}
Mosthaf, K., Baber, K., Flemisch, B., Helmig, R., Leijnse, A., Rybak, I.,
  Wohlmuth, B.: A new coupling concept for two-phase compositional porous media
  and single-phase compositional free flow, Water Resour. Res. 47, 1--19 (2011)

\bibitem{Saffman:1971}
Saffman, P.: On the boundary condition at the surface of a porous medium,
  Studies in Applied Mathematics 50 (1971)

\bibitem{OchoaTapia:1995}
Ochoa-Tapia, J., Whitaker, S.: Momentum transfer at the boundary between a
  porous medium and a homogeneous fluid—ii. comparison with experiment,
  International Journal of Heat and Mass Transfer 38, 2647 -- 2655 (1995)

\bibitem{Martys:1994}
Martys, N., Bentz, D.P., Garboczi, E.J.: Computer simulation study of the
  effective viscosity in brinkman's equation, Physics of Fluids 6, 1434--1439
  (1994)

\bibitem{Lundgren:1972}
Lundgren, T.S.: Slow flow through stationary random beds and suspensions of
  spheres, Journal of Fluid Mechanics 51, 273--299 (1972)

\bibitem{Zhang:2009}
Zhang, Q., Prospretti, A.: Pressure-driven flow in a two-dimensional channel
  with porous walls, Journal of Fluid Mechanics 631, 1--21 (2009)

\bibitem{Nabovati:2009}
Nabovati, A., Amon, C.: Hydrodynamic boundary condition at open-porous
  interface: A pore-level lattice {B}oltzmann study, Transport in Porous Media
  96, 83--95 (2013)

\bibitem{LIU:2011}
Liu, Q., Prospretti, A.: Pressure-driven flow in a channel with porous walls,
  Journal of Fluid Mechanics 679, 77--100 (2011)

\bibitem{Preclik:2015}
Preclik, T., Ruede, U.: Ultrascale simulations of non-smooth granular dynamics,
  Computational Particle Mechanics pp. 1--24 (2015)

\bibitem{feichtinger:2009}
Feichtinger, C., G{\"o}tz, J., Donath, S., Iglberger, K., R{\"u}de, U.:
  Walberla: Exploiting massively parallel systems for lattice {B}oltzmann
  simulations, in: Parallel Computing, pp. 241--260, Springer (2009)

\bibitem{Succi:1989}
Succi, S., Foti, E., Higuera, F.: Three-dimensional flows in complex geometries
  with the lattice boltzmann method, EPL (Europhysics Letters) 10, 433 (1989)

\bibitem{Singh:2000}
Singh, M., Mohanty, K.: Permeability of spatially correlated porous media,
  Chemical Engineering Science 55, 5393 -- 5403 (2000)

\bibitem{Bernsdorf:2000}
Bernsdorf, J., Brenner, G., Durst, F.: Numerical analysis of the pressure drop
  in porous media flow with lattice boltzmann {(BGK)} automata, Computer
  Physics Communications 129, 247 -- 255 (2000)

\bibitem{Kim:2001}
Kim, J., Lee, J., Lee, K.C.: Nonlinear correction to darcy's law for a flow
  through periodic arrays of elliptic cylinders, Physica A: Statistical
  Mechanics and its Applications 293, 13 -- 20 (2001)

\bibitem{Spaid:1997}
Spaid, M.A.A., Phelan, F.R.: Lattice {B}oltzmann methods for modeling
  microscale flow in fibrous porous media, Physics of Fluids 9, 2468--2474
  (1997)

\bibitem{Freed:1998}
Freed, D.M.: Lattice-{B}oltzmann method for macroscopic porous media modeling,
  International Journal of Modern Physics C 09, 1491--1503 (1998)

\bibitem{Martys:2001}
Martys, N.S.: Improved approximation of the brinkman equation using a lattice
  {B}oltzmann method, Physics of Fluids 13, 1807--1810 (2001)

\bibitem{Nithiarasu:1997}
Nithiarasu, P., Seetharamu, K., Sundararajan, T.: Natural convective heat
  transfer in a fluid saturated variable porosity medium, International Journal
  of Heat and Mass Transfer 40, 3955 -- 3967 (1997)

\bibitem{Guo:2002}
Guo, Z., Zhao, T.: Lattice boltzmann model for incompressible flows through
  porous media, Physical Review E 66, 036304 (2002)

\bibitem{Bhatnagar:1954}
Bhatnagar, P.L., Gross, E.P., Krook, M.: A model for collision processes in
  gases. i. small amplitude processes in charged and neutral one-component
  systems, Phys. Rev. 94, 511--525 (1954)

\bibitem{Pan:2006}
Pan, C., Luo, L.S., Miller, C.T.: An evaluation of lattice {B}oltzmann schemes
  for porous medium flow simulation, Computers and Fluids 35, 898 -- 909 (2006)

\bibitem{Bogner2015}
Bogner, S., Mohanty, S., R{\"u}de, U.: Drag correlation for dilute and
  moderately dense fluid-particle systems using the lattice {B}oltzmann method,
  International Journal of Multiphase Flow 68, 71 -- 79 (2015)

\bibitem{Ginzburg:2007}
Ginzburg, I.: Lattice {B}oltzmann modeling with discontinuous collision
  components: Hydrodynamic and advection-diffusion equations, Journal of
  Statistical Physics 126, 157--206 (2007)

\bibitem{Ginzburg:2008:a}
Ginzburg, I., Verhaeghe, F., d'Humieres, D.: {Two-Relaxation-Time Lattice
  {B}oltzmann Scheme: About Parametrization, Velocity, Pressure and Mixed
  Boundary Conditions}, Commun. Comput. Phys. 3, 427--478 (2008)

\bibitem{Ginzburg:2008:b}
Ginzburg, I., Verhaeghe, F., d'Humi\`{e}res, D.: {Study of simple hydrodynamic
  solutions with the two-relaxation-times lattice-{B}oltzmann scheme},
  Communications in Computational Physics 3, 519--581 (2008)

\bibitem{He:1997}
He, X., Luo, L.S.: Lattice boltzmann model for the incompressible
  navier–stokes equation, Journal of Statistical Physics 88, 927--944 (1997)

\bibitem{Khirevich:2015}
Khirevich, S., Ginzburg, I., Tallarek, U.: Coarse-and fine-grid numerical
  behavior of {MRT/TRT} lattice {B}oltzmann schemes in regular and random
  sphere packings, Journal of Computational Physics 281, 708--742 (2015)

\bibitem{Feichtinger:2011}
Feichtinger, C., Donath, S., K\"ostler, H., G\"otz, J., R\"ude, U.: {WaLBerla:
  HPC software design for computational engineering simulations}, Journal of
  Computational Science 2, 105 -- 112 (2011)

\bibitem{Godenschwager:2013}
Godenschwager, C., Schornbaum, F., Bauer, M., K\"{o}stler, H., R\"{u}de, U.: A
  framework for hybrid parallel flow simulations with a trillion cells in
  complex geometries, in: Proceedings of SC13: International Conference for
  High Performance Computing, Networking, Storage and Analysis, SC '13, pp.
  35:1--35:12, ACM, New York, NY, USA (2013)

\bibitem{Peters:2010}
Peters, A., Melchionna, S., Kaxiras, E., L\"att, J., Sircar, J., Bernaschi, M.,
  Bison, M., Succi, S.: Multiscale simulation of cardiovascular flows on the
  {IBM Bluegene/P}: Full heart-circulation system at red-blood cell resolution,
  in: Proceedings of the 2010 {ACM/IEEE} International Conference for High
  Performance Computing, Networking, Storage and Analysis, pp. 1--10, IEEE
  Computer Society (2010)

\bibitem{Krafczyk:2011}
Sch{\"{o}}nherr, M., Kucher, K., Geier, M., Stiebler, M., Freudiger, S.,
  Krafczyk, M.: Multi-thread implementations of the lattice {B}oltzmann method
  on non-uniform grids for {CPUs} and {GPUs}, Computers and Mathematics with
  Applications 61, 3730--3743 (2011)

\bibitem{Robertsen:2015}
Robertsen, F., Westerholm, J., Mattila, K.: Lattice {B}oltzmann simulations at
  petascale on multi-{GPU} systems with asynchronous data transfer and strictly
  enforced memory read alignment, in: Parallel, Distributed and Network-Based
  Processing (PDP), 2015 23rd Euromicro International Conference on, pp.
  604--609 (2015)

\bibitem{Fattahi:2015}
Fattahi, E., Waluga, C., Wohlmuth, B., R\"ude, U., Helmig, R., Manhart, M.:
  Pore-scale lattice {B}oltzmann simulation of laminar and turbulent flow
  through a sphere pack, Submitted to Computers $\&$ Mathematics with
  Applications  (2015)

\end{thebibliography}
\end{document}